\documentclass[11pt,a4paper]{article}

\usepackage{amsmath,amssymb}
\usepackage{epsfig,graphicx}
\usepackage{subfigure}
\usepackage{graphicx}
\usepackage{rotating}
\usepackage{cancel}
\usepackage{bm}
\usepackage{color}
\usepackage{comment}
\usepackage{cite}

\renewcommand{\Re}{\mathrm{Re}}
\renewcommand{\Im}{\mathrm{Im}}

\renewcommand\[{\left[}

\newcommand{\exclude}[1]{}

\def\beq{\begin{equation}}
\def\eeq{\end{equation}}

\begin{document}

\numberwithin{equation}{section}
\title{{\normalsize  \mbox{}\hfill IPPP/14/99, DCPT/14/198}\\
\vspace{2.5cm} 
\Large{\textbf{An upper limit on the scale of new physics phenomena\\[0.1cm]
\normalsize{from\\
rising cross sections in high multiplicity Higgs and vector boson events} \vspace{0.5cm}}}}

\author{Joerg Jaeckel$^{1,a}$ and Valentin V. Khoze$^{2,b}$\\[2ex]
\small{\em $^1$Institut f\"ur theoretische Physik, Universit\"at Heidelberg,}\\
\small{\em Philosophenweg 16, 69120 Heidelberg, Germany}\\[0.5ex]
  \small{\em $^2$Institute for Particle Physics Phenomenology, Department of Physics,} \\
  \small{\em South Road, Durham DH1 3LE, United Kingdom}\\[0.8ex]
}

\date{}
\maketitle
\thispagestyle{empty}
\begin{abstract}
\noindent
In very high energy scattering events, production of multiple Higgs and electroweak gauge bosons becomes possible.
Indeed the perturbative cross section for these processes grows with increasing energy, eventually violating perturbative unitarity. 
In addition to perturbative unitarity we also examine constraints on high multiplicity processes arising from experimentally measured quantities.
These include the shape of the $Z$-peak and upper limits on scattering cross sections of cosmic rays.
We find that the rate of high multiplicity electroweak processes will exceed these upper limits at energies not significantly above what can be currently tested experimentally.
This leaves two options: 1) The electroweak sector becomes truly non-perturbative in this regime or 2) Additional physics beyond the Standard Model is needed. In both cases novel physics phenomena must set in before these energies are reached. Based on the measured Higgs mass we estimate the critical energy to be in the range of $10^3$~TeV but we also point out that it can potentially be significantly less than that.
\noindent
\end{abstract}

\vspace*{4cm}
\footnoterule
\vspace*{0.1cm}
\noindent
$^{a}${\footnotesize jjaeckel@thphys.uni-heidelberg.de}; $^{b}${\footnotesize valya.khoze@durham.ac.uk}

\newpage
\setcounter{page}{1}
\section{Introduction}
Already before turning on the LHC we knew we would find something new. Famously the argument~\cite{Lee:1977yc} was that either there is a Higgs boson with a mass below $\sim1$~TeV, new physics beyond the Standard Model, or scattering processes between electroweak gauge bosons becomes non-perturbative. One of the three options had to be realised because otherwise perturbative cross sections for $VV\to VV$ scattering would violate unitarity.
Indeed this question has been answered by the observation of a Higgs at $125$~GeV.

Of course one can now ask, whether the Standard Model including the Higgs is valid and perturbative up to arbitrarily high energies.
The measured Higgs mass of 125~GeV already points to the possibility that the Higgs potential becomes meta-stable at a scale of about $10^{10}$~GeV, unless new stabilising effects appear~\cite{stability}.
However, this scale is unobtainable in particle collider experiments.

Building on previous results~\cite{Cornwall:1990hh,Goldberg:1990qk,Voloshin:1992mz,Argyres:1992np,Brown:1992ay,Voloshin:1992rr,Libanov:1994ug,Son:1995wz,Khoze:2014zha,Khoze:2014kka} (for reviews see \cite{Voloshin:1994yp,Libanov:1997nt}) we will argue in this paper that perturbative unitarity, now for $H\to nH+mV$ or $V\to nH+mV$ production with $n+m\ggg 1$, provides once again a limit on the onset of new phenomena -- either non-perturbative behaviour or new physics beyond the Standard Model. 
Using the measured value of the Higgs mass we find that this must happen at energies lower than $1570$~TeV for a very conservative estimate of the 
cross section, and at energies $\lesssim 812$~TeV for a more realistic one.
We will also show how with a more speculative and optimistic interpretation, the limit might be lowered to $\lesssim 35$~TeV.

The standard unitarity and perturbativity arguments are formal statements which involve unknown, possibly large constants. 
Therefore we also consider two more physical arguments. 
The first considers the spectral representation of the propagator in terms of $1\to n$ matrix elements.
Considering the experimentally tested and testable low energy behaviour of the propagator we can infer a limit on the maximal size of matrix elements.
An even more direct bound can be obtained by looking at the cross section of the process $VV\to nH+mV$ with both $V$ on-shell. This is a proper physical scattering process. We argue that this cross section rises rapidly with energy and at a reasonably low energy even exceeds the size of the Universe. Now, if any combination of suitable particles, such as protons, exceeds the required centre of mass energy, it would immediately scatter and be turned into many Higgses and vector bosons with lower energies, thereby providing an effective limit on the maximal energy of cosmic rays which is not observed. We slightly refine this argument based on the observed flux of cosmic rays and find the scale for new phenomena to be $\lesssim 830$~TeV ($\lesssim1590$~TeV for a more conservative estimate of the cross section).

All these estimates are based on relatively conservative approximations for the cross sections. As we discuss in the conclusions the onset of new phenomena can occur considerably earlier.

This note is structured as follows. In Section~\ref{unitarity} we discuss limits on rising scattering amplitudes and cross sections from unitarity, perturbativity,
the low energy behaviour of the propagator and from cosmic ray physics. In Section~\ref{scale} we apply these limits in a somewhat naive manner to derive a first estimate of the scale at which new physical phenomena must occur. To do this we consider the production of a large number of Higgses but also combinations of Higgs and vector bosons. We turn to more proper physical scattering processes needed for the limit based on cosmic rays in Section~\ref{physical}.  We conclude in Section~\ref{conclusions} with a discussion on possible implications for observation and experiment.

\section{Limits on large and rising amplitudes and cross sections}\label{unitarity}
\subsection{Unitarity and perturbativity arguments}\label{unitaritysub}

For any process unitarity is nothing but the statement that the probability of something happening should not exceed $1$,
\begin{equation}
\label{matrixlimit}
1\geq\sum_{n\neq a}|M(a\to n)|^2.
\end{equation}
In other words the sum over all matrix elements squared should not exceed $1$.

For the matrix elements this can be seen from the optical theorem (which follows directly from the unitarity of the $S$-matrix and the definition of the matrix elements),
\begin{equation}
\label{optical}
2\Im[M(a\to a)]=\sum_{n} |M(a\to n)|^{2}=|M(a\to a)|^2+\sum_{n\neq a}|M(a\to n)|^2.
\end{equation}
This can be rewritten to read
\begin{equation}
-(\Im[M(a\to a)]-1)^2-(\Re[M(a\to a)])^2+1=\sum_{n\neq a}|M(a\to n)|^2
\end{equation}
which leads to the limit Eq.~\eqref{matrixlimit}.

For scattering processes one could be tempted to equate the right hand side with the phase space integrated matrix elements for 
$n$ particle final states, starting from an $a$ particle state,
\begin{equation}
\sum_{n}\int d\Pi |M|^2=\sum_{n}\left(\prod_{i=1}^{n}\int \frac{d^{3}p_{i}}{(2\pi)^3 E_{i}}\right)|M(k_{1},\ldots,k_{a}\to p_{1},\ldots,p_{n})|^2 (2\pi)^4\delta\left(\sum k_{i} -\sum p_{i}\right)
\end{equation}
One could now naively apply the limit Eq.~\eqref{matrixlimit} to this expression. We will call this ``naive unitarity'' limit,
\begin{equation}
\label{naive}
\sum_{n}\int d\Pi |M|^2\leq 1.
\end{equation}

However for infinite volumes and times there are normalisation issues and the naive unitarity limit is not strict.
To see this let us consider the scattering of $2$ particles.
The elastic part can be decomposed into partial waves corresponding to angular momentum eigenstates, it then reads
\begin{equation}
M_{\rm elastic}(2\to 2)=\sum_{l} 16\pi(2l+1)\,a_{l}\,P_{l}(\cos(\theta)),
\end{equation}
where $P_{l}$ are Legendre polynomials and $a_{l}$ is the partial wave amplitude.

We can now insert this into Eq.~\eqref{optical}.
On the left hand side of the Eq.~\eqref{optical} we have the amplitude for no change to the state at all, therefore $\theta=0$. On the right hand side
for the elastic part we can integrate over the scattering angle. This yields,
\begin{eqnarray}
2\,\Im\left[16\pi \sum_{l} (2l+1) a_{l}\right]\!\!&=&\!\!32\pi\sum_{l}(2l+1)|a_{l}|^2
+\sum_{n,{\rm inelastic}} 
\int d\Pi |M|^2.
\end{eqnarray}
The sum over $l$ has infinitely many terms. Therefore the simple argument does not work. This can be seen by bringing all elastic terms to the right hand side,
\begin{eqnarray}
32\pi\sum_{l}(2l+1)\left[-\left(\Im\, a_{l}-\frac{1}{2}\right)^{2}-\left(\Re \,a_{l}\right)^2+\frac{1}{4}\right]
=\sum_{n,{\rm inelastic}} 
\int d\Pi |M|^2.
\end{eqnarray}
Each partial wave can now ideally contribute $+1/4$ and the sum can reach arbitrary values. Accordingly the inelastic part is not directly constrained.
Nevertheless, if we could argue that only a finite number of partial $l_{\rm max}$ waves contributes, we would have
\begin{equation}
\label{partiallimit}
\sum_{n,{\rm inelastic}}\int d\Pi |M|^2\leq8\pi(l_{\rm max}+1)^2.
\end{equation}
Up to a (potentially large or even infinite) factor $8\pi(l_{\rm max}+1)^2$, this is the naive unitarity bound.

An alternative argument can be made that at least perturbativity breaks down in some sense. 
The cross section for a typical tree-level $2\to2$ scattering process is
\begin{equation}
\sigma_{\rm typ}\sim \frac{g^2}{4\pi}\frac{1}{s},
\end{equation}
with some coupling $g$ which should be small.
Roughly speaking this scattering becomes non-perturbative for 
\begin{equation}
g\sim 4\pi.
\end{equation}
We can now compare this to the total inelastic cross section,
\begin{equation}
\sigma_{\rm inelastic}=\frac{1}{2s}\sum_{n,{\rm inelastic}}\int d\Pi |M|^2.
\end{equation}
They are of similar size for 
\begin{equation}
\label{bound213}
\sum_{n,{\rm inelastic}}\int d\Pi |M|^2
\sim 8\pi\,, \quad {\rm so\,\, that} \quad 
\sigma_{\rm inelastic} \,\sim\, \frac{4\pi}{s}.
\end{equation}
When the bound~\eqref{bound213} is saturated it is reasonable to expect that some non-perturbative behaviour sets in.

Another similar well motivated limit\footnote{We thank Gavin Salam
for pointing this out to us.}, can be written down for the $WW$ cross section in the Gauge-Higgs theory, in terms of the
`geometric' cross section for scattering of two vector bosons of the transverse size $\sim 1/M_W$,
\begin{equation}
\sigma_{WW}\, \lesssim\,  \frac{4\pi^2}{M_W^2}.
\end{equation}
Essentially vector bosons undergo weak scattering only when within a distance 
of $\sim 1/M_W$ from each other because their masses ensure that weak interactions are short range. When the predicted
cross section exceeds this bound, something non-trivial or non-perturbative should happen to either fix it or explain the effect.
In practical terms, both these perturbative unitarity limits will lead to similar energy upper bounds on the growing cross sections at high multiplicity.

\subsection{Low energy behaviour of the propagator and optimal truncation}\label{propagator}

The unitarity limits we have discussed so far are somewhat formal  in the sense that they are not directly linked with observation. 
Let us therefore look at more phenomenological arguments.

An observable quantity is the propagator in momentum space,
\begin{equation}
\Delta(p)=\int d^{4} x \exp(ipx) \left\langle 0|T\phi(x)\phi(0)|0\right\rangle=\int^{\infty}_{0} ds\, \frac{\rho(s)}{p^2-s},
\end{equation}
where we have used the K\"all\'en-Lehmann spectral representation, and $\rho(s)$ is the spectral density,
\begin{eqnarray}
\label{spectral}
\rho(s)&=&\sum_{n} \left |\left\langle 0|\phi | n\right \rangle \right|^2\delta\left(\sqrt{s}-\sum_{i=1}^{n} p_{i}\right)
\\\nonumber
&=&
\sum_{n}\int d\Pi |M(1\to n)|^{2}(s)
\,\,=\,\,Z\delta(s-m^2_{\phi})+ \sum_{n\geq 2}\int d\Pi |M(1\to n)|^{2}(s).
\end{eqnarray}
The right hand side therefore contains exactly the phase space integrated matrix elements for off-shell $1\to n$ processes with 
energy $\sqrt{s}$.
Plugging \eqref{spectral} into the equation for the propagator we obtain,
\begin{equation}
\label{spec-rep}
\Delta(p)=\frac{Z_\phi}{p^2-m^2_{\phi}}+\sum_{n\geq 2} \int^{\infty}_{(nm_{\phi})^2} ds\, \,\frac{\int d\Pi |M(1\to n)|^{2}(s)}{p^2-s}.
\end{equation}
For $|p^2|<4m^{2}_{\phi}$ the second term on the right hand side gives a non-singular contribution to the propagator; the residue of the propagator pole is entirely determined by the first term.

The probability rates for
$1\to n$ processes  thus appear in \eqref{spec-rep} as the higher order corrections to the propagator.
One can now argue~\cite{Libanov:1997nt} that to guarantee perturbativity, or more precisely asymptotic behaviour
of the perturbation series, the higher order terms in $n$ on the right hand side of  \eqref{spec-rep} should not be too large.
Indeed a common heuristic~\cite{Boyd} suggests that the optimal point for the truncation of an asymptotic series is at the order which gives the smallest contribution, i.e. just before the higher order terms start to grow (and ultimately diverge). In our case this would be at the point where for a given energy
the phase space integrated matrix elements on the right hand side of Eq.~\eqref{spectral} start to grow with $n$.
The lowest energy when this happens would give us a the scale where a non-perturbative repair-mechanism should set in.

However, if there is no repair-mechanism within the theory, there is no obvious reason why the new effects should set in at the ``optimal truncation point''. Indeed repair by new physics could set in earlier or later than that. 
Nevertheless, the cure must happen before it spoils current observations and experiments. So let us now turn to  such more direct limits.

\bigskip 

High multiplicity processes start to contribute only at high energies. 
Indeed a process $1\to n$ only sets in at an energy
$\sqrt{s}=nm_{\phi}$ with $m_{\phi}$ the mass of the produced particle. Let us now consider the contribution of such a process to the propagator,
\begin{eqnarray}
&&\!\!\!\!\!\!\!\!\int^{\infty}_{(nm_{\phi})^{2}}ds\,\frac{\int d\Pi |M(1\to n)|^2(s)}{p^2-s}
\\\nonumber
&&\qquad\qquad\qquad\qquad\approx -\frac{1}{m^2_{\phi}} \left[m^{2}_{\phi}\int^{\infty}_{(nm_{\phi})^{2}}ds\,\frac{\int d\Pi |M(1\to n)|^2(s)}{s}\right]
\\\nonumber
&&\qquad\qquad\qquad\qquad\qquad\qquad-\frac{p^2}{m^4_{\phi}} \left[m^4_{\phi}\int^{\infty}_{(nm_{\phi})^{2}}ds\,\frac{\int d\Pi |M(1\to n)|^2(s)}{s^2}\right] +\ldots.
\\\nonumber
&&\qquad\qquad\qquad\qquad=\frac{1}{m^{2}_{\phi}}{\mathcal{C}}_{1}+\frac{p^2}{m^{4}_{\phi}}{\mathcal{C}}_{2}+\ldots
\end{eqnarray}
On the right hand side we have expanded in powers of $p^2$ and this expansion should be reasonable as long as $p^2\ll (nm_{\phi})^2$.

For small momenta $p^2\sim m^{2}_{\phi}$ we should not observe strong deviations from the perturbative behaviour
\begin{equation}
\Delta(p)_{\rm perturbative}\sim \frac{1}{p^2-m^2_{\phi}}.
\end{equation}

Let us consider only the deviation from this behaviour caused by the $1\to n$ process for a given fixed $n$,
\begin{equation}
\Delta(p)= \Delta(p)_{\rm perturbative}+\frac{1}{m^{2}_{\phi}}{\mathcal{C}}_{1}+\frac{p^2}{m^{4}_{\phi}}{\mathcal{C}}_{2},\qquad {\rm for}\quad p^2\ll (nm_{\phi})^2.
\end{equation}
The last two terms on the right hand side should be small in order not to be in conflict with present experiments.

To quantify this, let us consider the case where $\phi$ is the $Z$-boson. Taking into account the $Z$-width $\Gamma_{Z}$ the relevant observable is essentially the absolute value of the propagator squared which should give us the Breit-Wigner shape of the $Z$-boson,
observed at 
LEP, 
\begin{equation}
\sigma(E) \sim \frac{1}{(E^2-m^2_{Z})^2+m^{2}_{Z}\Gamma^{2}_{Z}}+2\Re\left[\frac{\frac{1}{m^{2}_{Z}}{\mathcal{C}}_{1}+\frac{E^2}{m^{4}_{Z}}{\mathcal{C}}_{2}}{(m^2_{Z}-E^{2})+im_{Z}\Gamma_{Z}}\right]+\left(\frac{1}{m^{2}_{Z}}{\mathcal{C}}_{1}+\frac{E^2}{m^{4}_{Z}}{\mathcal{C}}_{2}\right)^2.
\end{equation}
Where $\sigma(E)$ is, e.g., the cross section for $e^{+}e^{-}\to hadrons$ near the $Z$-peak.

\begin{figure}[t]
\centering
   \includegraphics[width=6cm]{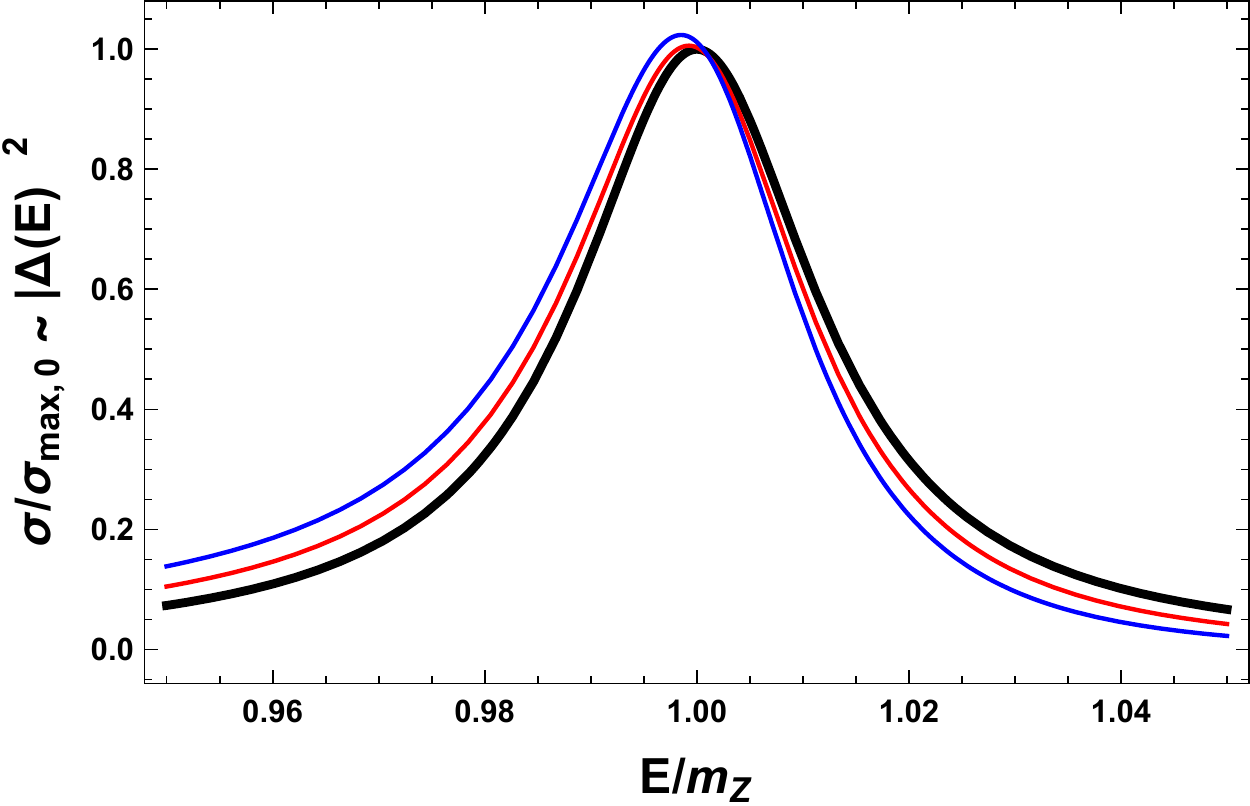} 
\hspace*{2cm}
   \includegraphics[width=6cm]{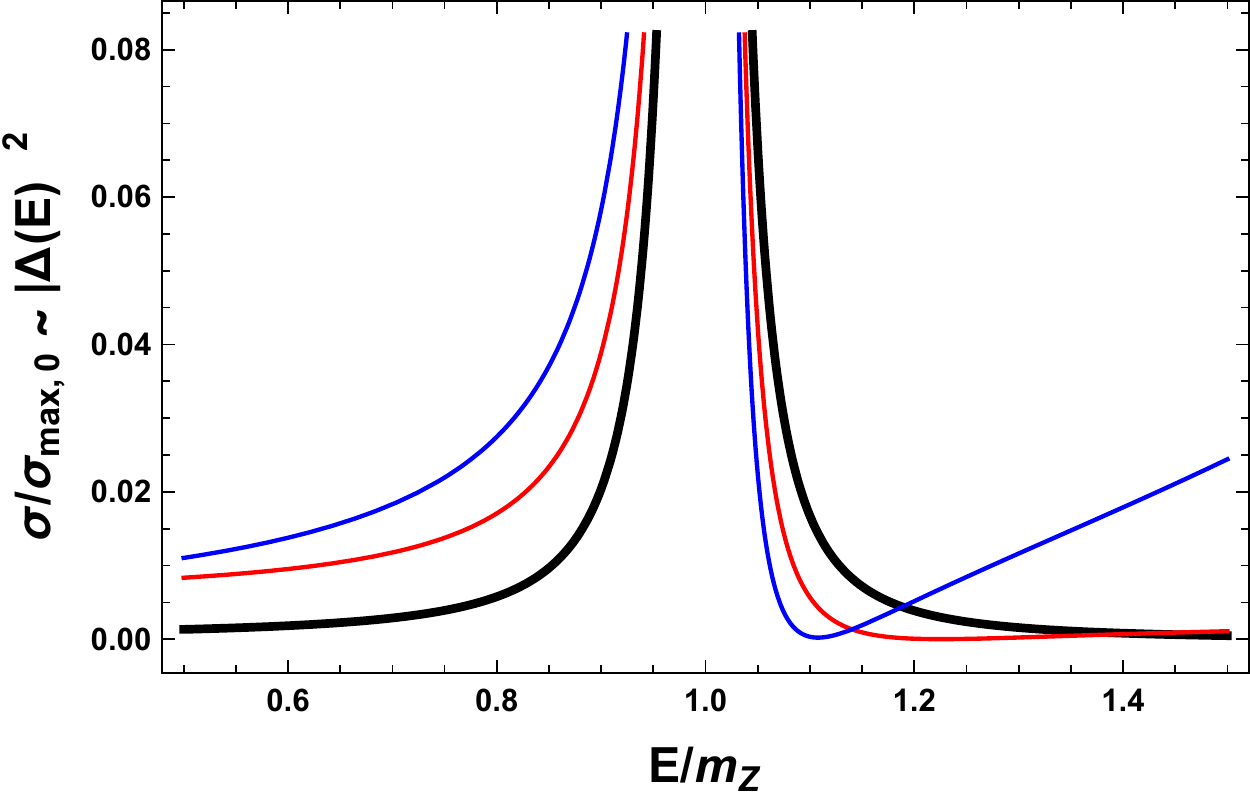} 
  \caption{The Breit-Wigner form of the $Z$-peak (normalized to 1) and modifications caused by unsuppressed high multiplicity processes. The black curve shows
   the standard Breit-Wigner shape, red corresponds to ${\mathcal{C}}_{1}=2,{\mathcal{C}}_{2}=0$ and blue
   to ${\mathcal{C}}_{1}=2,{\mathcal{C}}_{2}=2$. The left panel shows the immediate vicinity of the $Z$-mass. A larger region is shown in the right panel.
   Such large deviations are excluded by observation~\cite{ALEPH:2005ab}.}
   \label{Zfigure}
\end{figure}

In Fig.~\ref{Zfigure} we show the deviations caused to the Breit-Wigner shape for different ${\mathcal{C}_{1}}\sim{\mathcal{C}_{2}}\sim 1$.
Such strong deviations are in conflict with the experiment~\cite{ALEPH:2005ab}.
Accordingly we can limit
\begin{eqnarray}
\left|{\mathcal{C}}_{1}\right|&=&\left|m^2_{\phi}\int^{\infty}_{(nm_{\phi})^{2}}ds\,\frac{\int d\Pi |M(1\to n)|^2(s)}{s}\right|\lesssim 1
\\\nonumber
\left|{\mathcal{C}}_{2}\right|&=&\left|m^4_{\phi}\int^{\infty}_{(nm_{\phi})^{2}}ds\,\frac{\int d\Pi |M(1\to n)|^2(s)}{s^2}\right|\lesssim 1.
\end{eqnarray} 

\subsection{Cosmic ray limit}\label{cosmicsub}
The considerations above based on the shape of the resonance have provided indirect constraints on the high-multiplicity processes
which in this context appear as higher order corrections.
Yet, if we consider direct tree-level scattering with high multiplicity --  $2\to n$  -- events, they are the leading order contribution.
We would like to obtain a direct limit on the cross sections and consequently on the matrix elements for such processes.

Observations tell us that there is a significant flux of cosmic rays arriving at Earth from essentially all possible directions. 
The main components are typically protons and heavier nuclei.

Now, let us assume that the effective cross section for the inelastic scattering of two cosmic rays is of the size of the Universe. 
Then two cosmic rays flying in opposite directions could never pass each other without undergoing an inelastic scattering, thereby loosing a significant amount of energy. This severely limits the distance a cosmic ray can travel. 
Even if there was only one cosmic ray per year flying in a certain direction, a cosmic ray travelling in 
the opposite direction
would typically scatter every year, thereby loosing its energy.
In consequence high energy cosmic rays would be severely attenuated in conflict with observation.

More quantitatively the flux of cosmic rays is usually given as,
\begin{equation}
{\mathcal{F}}=\frac{dN}{dt dA d\Omega dE}.
\end{equation}
I.e. it is the number of cosmic rays $dN$, per time $dt$, per area $dA$ and per spatial angle $d\Omega$.

\begin{figure}[t]
\centering
   \includegraphics[width=6cm]{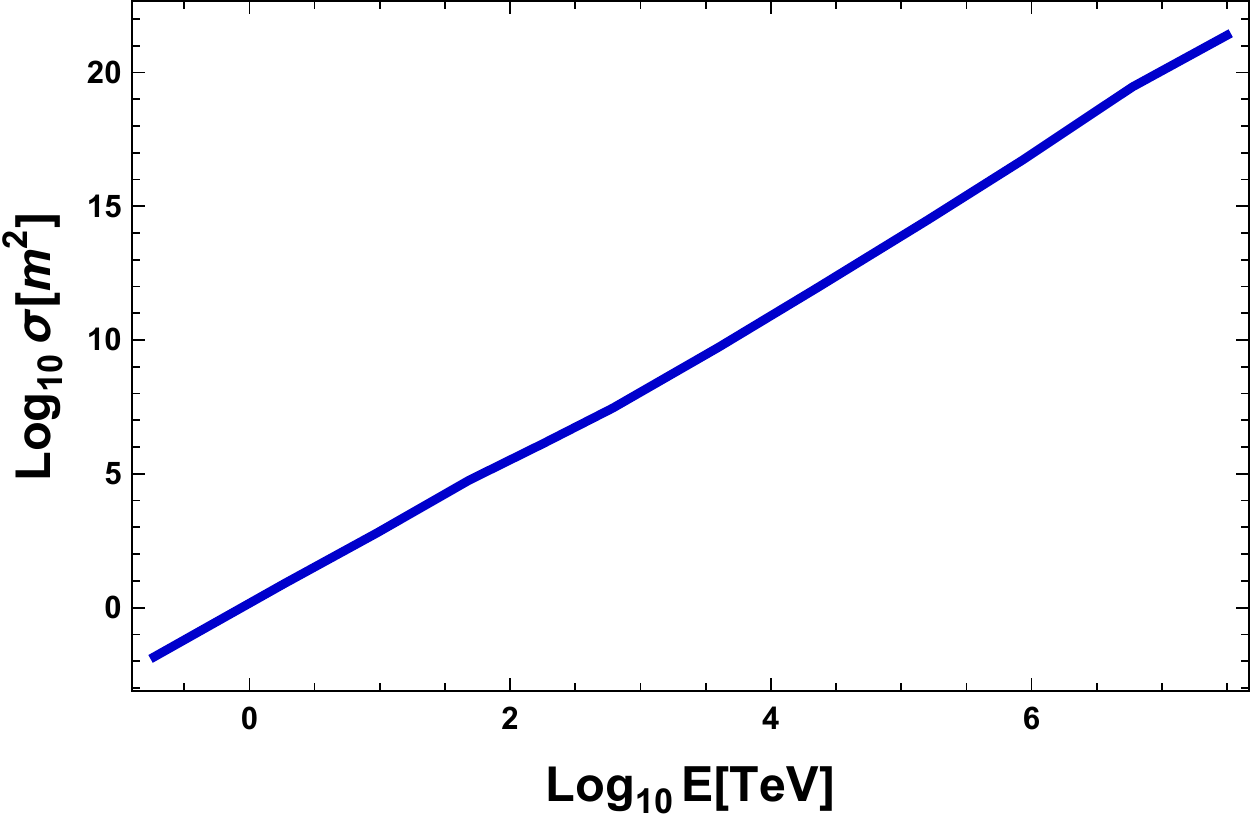} 
 \hspace*{2cm}  \includegraphics[width=6cm]{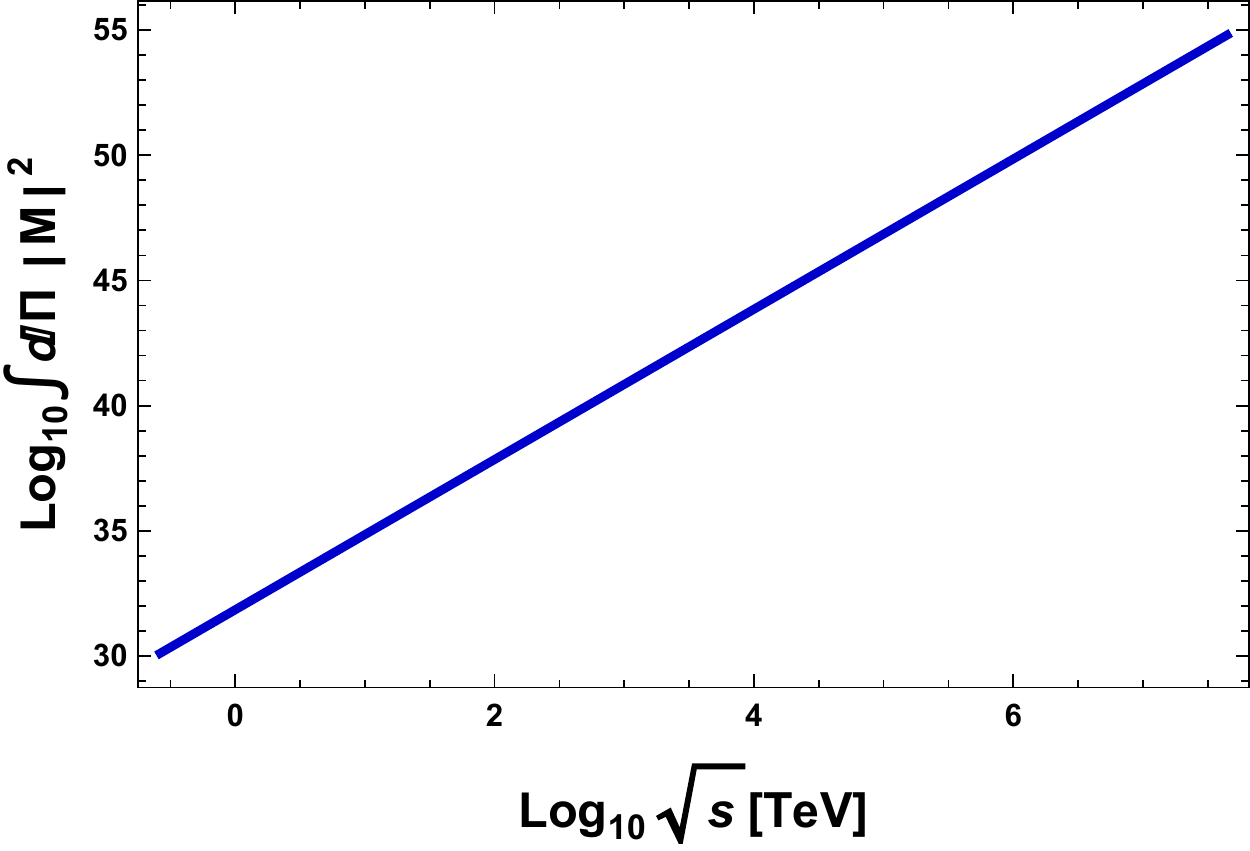} 
   \caption{Limits on the cross section and phase space integrated matrix element from the observation of cosmic rays. Assuming a minimal source distance of incoming cosmic rays of $\gtrsim 1$~lightyear.}
   \label{cosmicfig}
\end{figure}

A cosmic ray travelling in a fixed direction now sees an incoming density of cosmic rays flying in essentially the opposite direction and with roughly the same energy,
\begin{equation}
\rho_{\rm incoming}\sim {\mathcal{F}} \Delta\Omega\Delta E.
\end{equation}
Accordingly the mean free path is,
\begin{equation}
\ell_{\rm MFP}\sim\frac{1}{\sigma(E)\rho_{\rm incoming}}=\frac{1}{\sigma(E)}\frac{1}{{\mathcal{F}}\Delta\Omega\Delta E},
\end{equation}
and cosmic rays sourced at distances greater than $\ell_{\rm MFP}$ would be severely attenuated.
We can therefore limit the maximal allowed cross section for a given energy $E$ as,
\begin{equation}
\sigma(E)\lesssim \frac{1}{\ell_{\rm MFP}{\mathcal{F}}\Delta E\Delta\Omega}.
\end{equation}
Using the measured flux of cosmic rays (cf, e.g.~\cite{Nagano:2000ve}) and very conservative values for $\Delta E
\sim1$~GeV and $\Delta\Omega\sim 10^{-4}$~sr as well as $\ell_{\rm MFP}\sim 1$~lightyear we find the limit on the cross sections shown in the left panel of Fig.~\ref{cosmicfig}.

With
\begin{equation}
\sum_{n,{\rm inelastic}}\int d\Pi\, |M|^2\,=\,2s\,\sigma_{\rm inelastic}
\end{equation}
we can translate this into the limits on the integrated matrix elements shown in the right panel of Fig.~\ref{cosmicfig}.

This limit on the matrix elements is many orders of magnitude weaker than the naive 
perturbative unitarity limit. But as we will see below, the cross sections for high multiplicity Higgs and vector boson production rise so rapidly, that this has actually only a relatively small impact on the 
energy scale at which new phenomena must occur.

Our limit from cosmic rays is strictly speaking a limit on the $2\to n$ matrix elements, for on-shell initial states.
In the following section we will ignore the limitation, looking at simpler $1\to n$ matrix elements. We will return with more details on the $2\to n$ case in Section~\ref{physical}.

\section{The scale when new physical phenomena must occur}\label{scale}
Let us now apply the limits derived in the previous section to matrix elements for multiple Higgs and vector boson production.
It has already been noted in~~\cite{Cornwall:1990hh,Goldberg:1990qk,Voloshin:1992mz,Argyres:1992np,Brown:1992ay,Voloshin:1992rr,Libanov:1994ug,Son:1995wz,Khoze:2014zha,Khoze:2014kka} that these matrix elements grow rapidly signalling a breakdown in perturbation theory. Indeed as argued above some form of new phenomenon -- either non-perturbative behaviour, or the appearance of physics beyond the Standard Model -- must occur before the integrated matrix exceed their limits.

The simplest case is the matrix element for $H\to nH$ processes (where the initial $H$ is off-shell). 
At tree level one can obtain this amplitude purely from scalar field theory.
In~\cite{Voloshin:1992rr} a lower limit on the phase space integrated matrix element squared for this process has been obtained,
\begin{equation}
\label{voloshin}
\int d\Pi |M|^2\gtrsim \exp\left(\frac{32\pi^2}{3\lambda}f\right).
\end{equation}
Where the function
\begin{equation}
f=\alpha \nu\left[\log(\alpha)+\log(\nu)-1+\log\left[\int^{\omega}_{1}dz\,\exp(-z\tau)\sqrt{z^2-1}\right]-2\log(\omega)\right]+\alpha\tau.
\end{equation}
has to be evaluated at $\tau$ and $\omega$ which solve,
\begin{eqnarray}
\nu&=&\frac{\int^{\omega}_{1}dz\,\exp(-z\tau)\sqrt{z^2-1}}{\int^{\omega}_{1}dz\,\exp(-z\tau)z\sqrt{z^2-1}},
\\\nonumber
0&=&\int^{\omega}_{1}dz\,\exp(-z\tau)\sqrt{z^2-1}-\frac{1}{2}\exp(-\omega\tau)\omega\sqrt{\omega^2-1}.
\end{eqnarray}
Moreover we have,
\begin{equation}
\alpha=\frac{E}{m_{h}}\frac{3\lambda}{32\pi^2},\qquad\nu=\frac{nm_{h}}{E}.
\end{equation}

Numerically one can determine that $f$ exceeds zero for a suitable choice of $\nu$ for $\alpha=15.4$ and it exceeds
$40\,\log[10]\times 3\lambda/(32\pi^2)$ for the slightly larger value of $\alpha=15.6$. According to Eq.~\eqref{voloshin}
at these points the phase space integrated matrix elements exceed the naive unitarity limit $\sim 1$ and the cosmic ray limit $\sim 10^{40}$, respectively.
This corresponds to required energies of
\begin{eqnarray}
E&\lesssim&1570~{\rm TeV}\qquad {\rm naive\,\,unitarity\,\,limit}
\\\nonumber
E&\lesssim&1590~{\rm TeV}\qquad {\rm cosmic\,\,ray \,\,limit}.
\end{eqnarray}
The naive unitarity limit is in agreement with the estimate performed in~\cite{Voloshin:1992rr}. 

Note that the rapid rise in cross section with energy ensures that there is little difference between the naive estimate and the observations based cosmic ray limit.
The maximal energy arising from constraints on the propagator lies in between the two.

As argued in the previous section there also exists a lower characteristic energy scale which marks the onset of non-perturbative behaviour 
according to the heuristic optimal truncation rule for asymptotic series.
This is determined as follows. 
At a fixed energy value one can check whether the $n+1$ particle process has larger rate than the $n$-particle process. 
The lowest energy when this happens determines the scale in question.
Using the estimate for the matrix elements as before we obtain,
\begin{equation}
E\lesssim 970~{\rm TeV}\qquad {\rm asymptotic\,\,series\,\,truncation \,\,heuristic}.
\end{equation}

\bigskip 

In~\cite{Son:1995wz} a better estimate for the phase space integrated matrix element in unbroken scalar $\phi^4$ has been obtained. This result has then been extended to the spontaneously broken gauge-Higgs system~\cite{Khoze:2014zha,Khoze:2014kka},
\begin{eqnarray}
\label{formula}
&&\!\!\!\!\!\!\int d\Pi |M|^2(H\to nH+mV) 
\\\nonumber
&&\quad\quad\sim\exp\bigg[2\log(\kappa^m d(n,m))+n\log\left(\frac{\lambda n}{4}\right)+m\log\left(\frac{\lambda m}{4}\right)
\\\nonumber
&&\quad\quad\quad\quad+\frac{n}{2}\left(3\log\left(\frac{\varepsilon_{h}}{3\pi}\right)+1\right)+\frac{m}{2}\left(3\log\left(\frac{\varepsilon_{V}}{3\pi}\right)+1\right)
-\frac{25}{12}n\varepsilon_{h}-3.15m\varepsilon_{V}\bigg].
\end{eqnarray}
Here, the $d(n,m)$ are coefficients, typically bigger than 1 that have to be determined by a algorithm similar to the one discussed in the next section and 
\begin{equation}
\label{kappa}
\kappa=\frac{g}{2\sqrt{2\lambda}}=\frac{m_{V}}{m_{h}}\sim 80/125\sim 0.65.
\end{equation}
$\varepsilon_{h}$ and $\varepsilon_{V}$ denote the average fraction of kinetic particle carried by the Higgs and vector bosons, respectively. They are related to the total energy via,
\begin{equation}
E=n(1+\varepsilon_{h})m_{h}+m(1+\varepsilon_{V})m_{V}.
\end{equation}

In particular for pure Higgs production we can use $d(n,0)=1$.
Using this a new estimate can be obtained by maximising the exponent for a fixed total energy and determining when the value at the maximum exceeds the limits set by the considerations of the previous section.
This yields the significantly lower values,
\begin{eqnarray}
\label{resultnum}
E& \lesssim&812~{\rm TeV}\qquad  {\rm naive\,\,unitarity\,\,limit}
\\\nonumber
E& \lesssim&830~{\rm TeV}\qquad {\rm cosmic\,\,limit}
\\\nonumber
E& \lesssim&299~{\rm TeV}\qquad {\rm asymptotic\,\,series\,\,truncation \,\,heuristic}.
\end{eqnarray}
We have checked, that unless there is a strong growth in the coefficients $d(n,m)$, the limit is not improved by considering the production of a large number of vector bosons.

The estimates~\eqref{resultnum} follow from using Eq.~\eqref{formula} which has been derived in the nonrelativistic limit where $\varepsilon$ is small. 
More precisely the expression for the cross section~\eqref{formula} is justified in the double scaling limit $\varepsilon\to 0$, $n\varepsilon={\rm fixed}$.
We have extrapolated these expressions to the regime of moderate $\varepsilon\simeq 1$ without taking into account unknown corrections of order $\varepsilon^2$
and beyond. There is an inherent sensitivity to higher orders in $\varepsilon$. 
Specifically, adding a term $\sim n\varepsilon^2$ with a coefficient of order $1$ in the exponent of the cross section formula can lower the energy scale for the onset of new phenomena down to below 100~TeV,
\begin{equation}
\label{res-neps2}
E \lesssim 100 ~{\rm TeV}\qquad  \sim n\varepsilon^2 \,\, {\rm factor\,\,effect.}
\end{equation}
We will further comment on this in the conclusions. 

Before drawing definitive conclusions from the tree-level formulae we have been using, one should also consider the effect 
of loop corrections\footnote{We would like to thank Martin Bauer and Tilman Plehn for discussions on this point.}. 
In the scalar sector of the theory it was argued in~\cite{Libanov:1994ug} that the leading one-loop correction to the high-multiplicity amplitude on the threshold, computed in \cite{Smith:1992rq},
\begin{equation}
 M(H\to nH)_{\rm tree + 1 loop}^{\rm thr.}\,=\, n!\, (2v)^{1-n} \left(1+n(n-1)\frac{\sqrt{3} \lambda}{8\pi}\right)
\,,
\label{eq:amplnh1}
\end{equation}
in fact exponentiates, so that we have,
\begin{equation}
|M|^2(H\to nH)^{\rm thr.} \,=\,|M|^2(H\to nH)_{\rm tree}\times\, \exp\left[\frac{\sqrt{3} }{4\pi}\, \lambda n^2\,+\, {\cal O}((\lambda n)^2)\right]\,.
\label{expB} 
\end{equation}
This expression holds in the limit $\lambda \to 0,$ $n\to \infty$ with $\lambda n^2$ being fixed.
Of course, the higher-order corrections $\sim n\lambda$ are important, as we are 
interested in multiplicities $n \sim 1/\lambda$, and
we should stress that this such exponential enhancement is only guaranteed at the leading order in $n^2 \lambda$. Furthermore
we have no control over the
momentum dependence of these loop corrections for amplitudes away from the multiparticle threshold. 
But taken at face value, \eqref{expB} leads to 
an exponential enhancement of our tree-level cross section formula \eqref{formula} by a factor $e^{\frac{\sqrt{3} }{4\pi}\, \lambda n^2}$.
Using such $\sim +\lambda n^2$ corrections to the exponent 
we  obtain the limit
\begin{equation}
\label{res-loop}
E \lesssim 35 ~{\rm TeV}\qquad  {\rm naive\,\,loop\,\,factor\,\,effect,}
\end{equation}
which seems very promising for applications at the FCC.
However, we stress again that in the relevant for us high-energy, high-multiplicity
regime the employed approximation is questionable and improved calculations are clearly needed.

\section{Physical $2\to n$ scattering processes}\label{physical}
Our discussion in the previous section had a slight deficiency. We were considering the phase space integrated matrix elements for an initial state of a single highly off-shell boson. In reality one should look at physical scattering processes which are $2\to n$ with two on-shell initial particles.
Indeed, the most convincing limit on the matrix elements arising from the cosmic ray argument directly relies on such $2\to n$ scattering processes.

In the simplest case of pure $\phi^4$ scalar field theory, both in the unbroken and in the broken phase, it is actually known that
tree-level amplitudes on the multiparticle threshold for $2\to n$ processes are exactly vanishing~\cite{Voloshin:1992nu,Smith:1992rq,Voloshin:1992xb,Voloshin:1992nm,Argyres:1992un}. Indeed some of this may even survive beyond tree-level~\cite{Voloshin:1992nm}.
If this were the case in the Standard Model, the conclusions drawn from the single particle state would loose their power.
However, pure scalar field theory is a very special case~\cite{Voloshin:1992xb,Argyres:1993xa,Smith:1993hz}.
In the Standard Model nullification of the threshold amplitudes in the two-particle scatterings occurs only for particular values of the ratio between the particle masses.

\begin{figure}[t]
\centering
 \includegraphics[width=10cm]{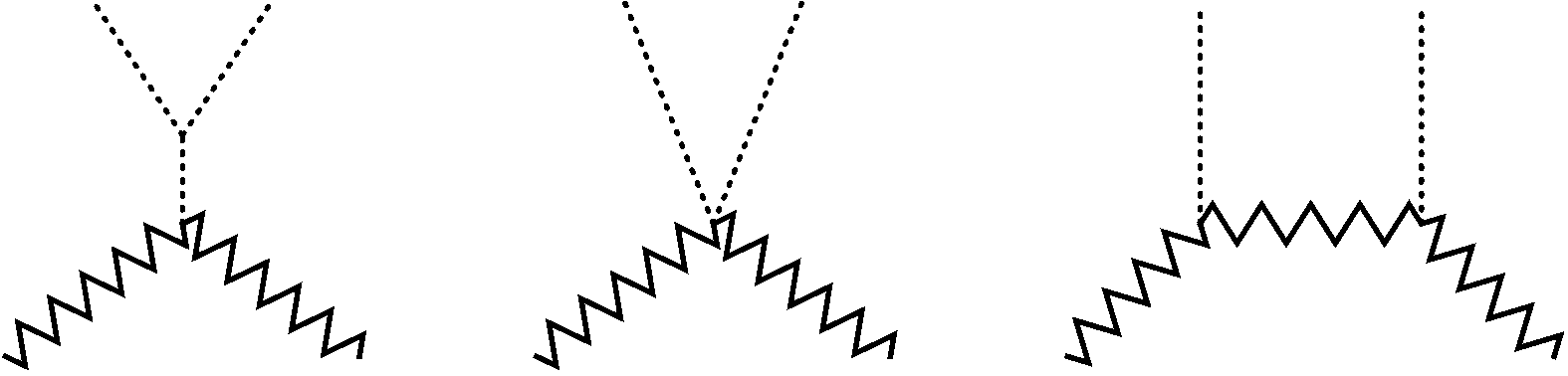} 
   \caption{Diagrams contributing to the $VV\to 2H$ process. The first diagram can be directly obtained by attaching a $H\to 2H$ diagram to the two vector bosons. It interferes however, with diagrams where the two Higgses are directly radiated off the vector bosons (second and third diagram)}
   \label{attach}
\end{figure}

We will now consider a specific Standard Model $2\to n$ process where two transverse on-shell vector bosons scatter to produce a large number of Higgs bosons. 
For pure multi-Higgs production one can upgrade the matrix element from $H\to nH$ to $VV\to nH$ simply by producing the off-shell Higgs from 
two initial vector bosons (first diagram in Fig.~\ref{attach}). However, the corresponding diagrams can then interfere with diagrams where multiple Higgses are
radiated off the vector bosons (second and third diagram in Fig.~\ref{attach}).
This interference could be destructive (we have negative t-channel propagators), which is exactly what happens in the pure scalar $\phi^4$ case.

Therefore let us extend the arguments used in~\cite{Brown:1992ay} to allow for a physical $2V$ initial state in order to take all relevant diagrams into account.
As explained in~\cite{Khoze:2014zha} (cf. Eqs.~(2.14) and (3.9)) the matrix element for the process $V\to V+nH$ with $n$ on-shell Higgses can be obtained from the solution to the classical equation of motion,
\begin{equation}
\label{simple}
d_{t}^2 A=-\frac{g^2}{4}h^2A.
\end{equation}
with 
\begin{equation}
h^{2}=v^2\left(\frac{1+\frac{z}{2v}}{1-\frac{z}{2v}}\right)^2=v^2\left(1+4\sum_{k=1}^{\infty}k\,z^{k}\right),\quad{\rm where}\quad z=z_{0}\exp(im_{h}t).
\end{equation}
We can now expand
\begin{equation}
A=\sum_{k} a_{n}\left(\frac{z}{2v}\right)^{n}.
\end{equation}
The matrix elements can be obtained by taking $n$ derivatives with respect to $z$ of this generating function $A$~\cite{Brown:1992ay,Khoze:2014zha},
\begin{equation}
\label{bla}
M=n!(2v)^{-n}a_{n}.
\end{equation}

To include a non-vanishing 3-momentum flowing through the vector bosons (the Higgses remain on-shell) we simply have to add the 
kinetic energy of the vector bosons to this equation (this is similar to the technique used in~\cite{Voloshin:1992xb}). This is most simple for the case of transverse gauge bosons in which case we have,
\begin{equation}
d^{2}_{t} A=-\left[\vec{p}^{\,2}+\frac{g^2}{4}h^2\right]A.
\end{equation}

When solving the equation we have to take into account that the second vector boson is actually an incoming particle and is also on-shell. It therefore contributes negatively to the energy.
We therefore have to solve
\begin{equation}
-(nm_{h}-E_{V})^2 A=-\left[\vec{p}^{\,2}+\frac{g^2}{4}h^2\right]A.
\end{equation}
with the on-shell conditions,
\begin{eqnarray}
E_{V}=\sqrt{\vec{p}^{\,2}+m^{2}_{V}}=\frac{n}{2}m_{h}.
\end{eqnarray}
The latter holds because the two vector bosons together have to provide the energy to produce the $n$ Higgses.

Comparing coefficients of $z^n$ on both sides of the equation we find the relation,
\begin{equation}
a_{l}=\frac{4\kappa^2}{l^2-2E_{V}l}\sum_{k=1}^{l-1}k\,a_{k}.
\end{equation}
In the last step when $l=n$ one encounters a divergence. This arises from the fact that the vector boson is then on-shell.
This is exactly the propagator one has to amputate when using the LSZ reduction formula. Accordingly in the last step one simply multiplies by $4\kappa^2$.

\begin{figure}[t]
\centering
   \includegraphics[width=6cm]{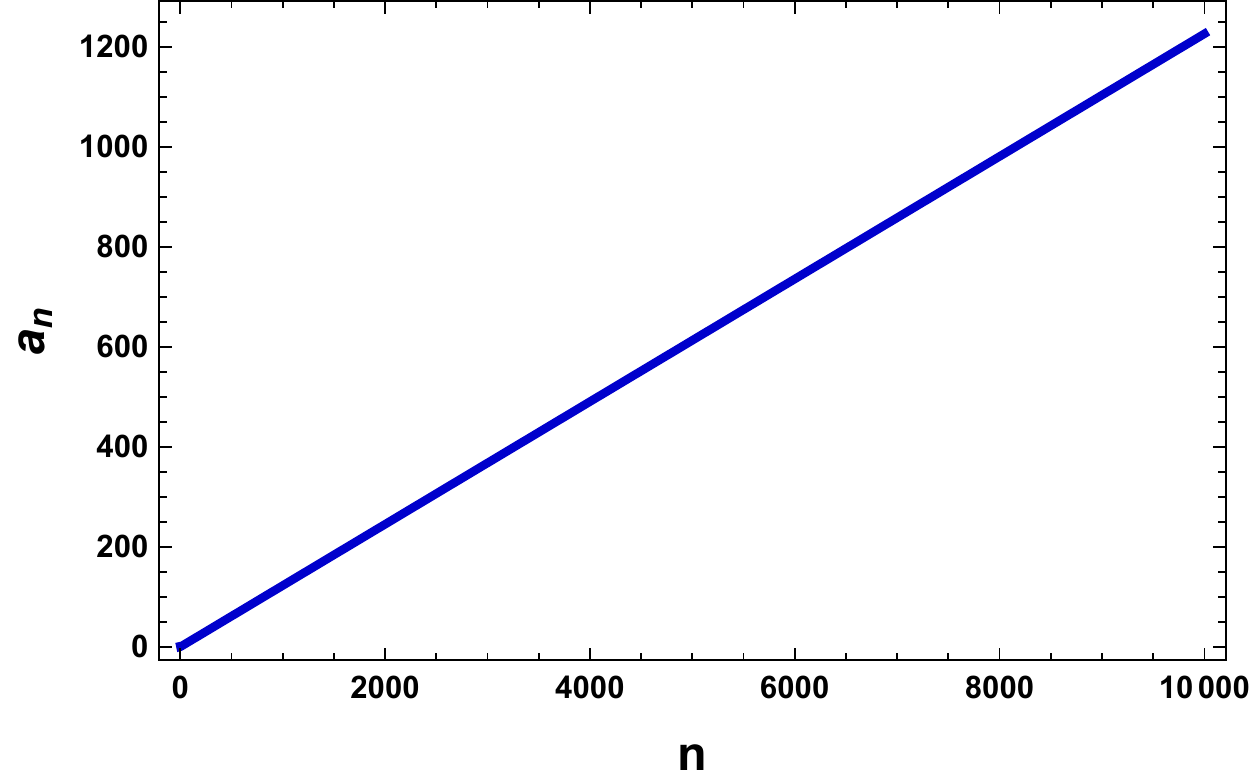} 
   \caption{Matrix element coefficients $a_{n}$ for $VV\to nH$.}
   \label{vaus}
\end{figure}

To facilitate fast computation this can be rewritten in terms of the recursion relations,
\begin{eqnarray}
b_{k}&=&b_{k-1}+c_{k-1}+a_{k-1}
\\\nonumber
c_{k}&=&c_{k-1}+a_{k-1}
\\\nonumber
a_{k}&=&\frac{4\kappa^2}{k^2-2E_{V}k}b_{k},
\end{eqnarray}
with the initial values,
\begin{equation}
b_{0}=0,\quad c_{0}=0,\quad a_{0}=1.
\end{equation}
In the last step one has to multiply $b_{n}$ only by $4\kappa^2$.
The resulting coefficients $a_{n}$ for a large range of values $n$ are shown in Fig.~\ref{vaus}.
We can see that with increasing $n$, $a_{n}$ grows approximately linearly. 
The dominant behaviour is therefore the factorial growth of the amplitude inherent in Eq.~\eqref{bla}. 
In consequence the estimates for the scale where new phenomena set in (obtained in the previous section), continue to hold in the more physical situation with two on-shell particles in the initial state.

Finally let us note that the dependence on $\kappa$ is quite strong.
Indeed for the value $\kappa=1/\sqrt{2}$ we have complete destructive interference in agreement with what has been observed in~\cite{Voloshin:1992nm}. 
We will discuss this further in the conclusions.

\section{Conclusions}\label{conclusions}
At very high energies it becomes energetically possible to produce multiple Higgs and vector bosons.
The naive expectation is that, as we increase the particle number these events become more and more unlikely because
perturbatively adding more particles in the final state means that we have to multiply by higher powers of the small coupling constant.
However, it has been shown~~\cite{Cornwall:1990hh,Goldberg:1990qk,Voloshin:1992mz,Argyres:1992np,Brown:1992ay,Voloshin:1992rr,Libanov:1994ug,Son:1995wz,Khoze:2014zha,Khoze:2014kka} that the amplitudes contributing to such processes grow factorially with the number of particles thereby overcoming any suppression of the small coupling, as long as the number of particles is large enough.
Therefore at some point the corresponding cross sections start to grow and can indeed become very large.
In this context the fundamental question is what this implies for the health of the SM and importantly also for potential observable effects.

Indeed rapidly growing cross sections are in conflict with arguments based on unitarity and perturbativity (Section~\ref{unitaritysub}). But as we have shown in this paper at an only slightly higher scale they also are in conflict with existing measurements of the $Z$-peak (Section~\ref{propagator}) 
and with observations of cosmic rays (Section~\ref{cosmicsub}). Therefore the perturbative treatment of the SM exhibits not simply a formal mathematical breakdown but will also be in conflict with observation.
We are therefore left with two options:
\begin{itemize}
\item{} At high energies (multiplicities) the SM is fundamentally non-perturbative.
\item{} New physics beyond the SM has to set in before the cross sections become too large.
\end{itemize}
In either case this requires new physics phenomena below the scale of the breakdown.
In this sense the situation is reminiscent of the unitarity bound on WW-scattering, that told us that either a Higgs, new physics or non-perturbative behaviour must be found below a scale of $\sim 1$~TeV.

For practical purposes there are two crucial questions. The first is at what scale do we expect new phenomena, the second is what one would actually observe.

Based on the measured Higgs mass and using a lower estimate on the phase space integrated matrix elements~\cite{Voloshin:1992rr}, one finds that new phenomena must set in at scales below $\lesssim 1600$~TeV. This is a very conservative upper bound. 
Using expressions for the cross sections obtained in the non-relativistic regime, we find a scale of $\lesssim 830$~TeV when cross sections exceed observational limits. Furthermore, already below that scale at energies of order $\sim 300$~TeV the cross sections exhibit peculiar behaviour in the sense that higher order/higher multiplicity processes are larger than lower multiplicity ones. It seems therefore quite plausible that new phenomena must set in before that scale.

The non-relativistic approximation may still be too pessimistic. In unbroken scalar $\phi^4$ theory, corrections to the non-relativistic approximation have been calculated and they tend to increase the cross section~\cite{Bezrukov:1995ta,Bezrukov:1995qh,Libanov:1997nt}. Adding a term $\sim n\varepsilon^2$ in the exponent of Eq.~\eqref{formula} with a coefficient $\simeq 1$ as motivated by results obtained in $\phi^4$ theory, and going to
the moderately relativistic regime the breakdown scale can be lowered below $100$~TeV. 
This would bring it into the region probed by a future high energy proton machine such as the FCC.
We have also considered the effects of loop corrections by adding the $\sim +\lambda n^2$ term to the exponent of the 
tree-level cross section which lowered the limit on the energy scale to below 35 TeV.
The summary of our energy upper bounds (using the cosmic limit on the matrix elements\footnote{Estimates based on the asymptotic series heuristic
are typically lower by a factor of a few.}) is as follows:
\begin{eqnarray}
E& \lesssim&1590~{\rm TeV}\qquad  {\rm upper\,\,limit\,\,from\,\,conservative\,\,1\to n\,\, kinematics}
\\ \nonumber
E& \lesssim&830~{\rm TeV}\qquad\,\,\, {\rm non-relativistic\,\,kinematics}
\\ \nonumber
E& \lesssim&100 ~{\rm TeV}\qquad \,\,\, {\rm include}\,\, \sim n\varepsilon^2 \,\, {\rm factor\,\,effect}
\\ \nonumber
E& \lesssim&\,\,35 ~{\rm TeV}\qquad  \,\,\,\,{\rm include\,\, naive\,\,loop\,\,factor\,\,effect.}
\end{eqnarray}
However, we stress that in deriving that last two numbers we employed some rather simplistic approximations.
Indeed, even before we switch on the $\sim n\varepsilon^2$ or loop-level corrections, we note that the full momentum dependence of tree-level
$2 \rightarrow n$ rates is not known, we have used instead the $1 \rightarrow n$ expressions.
Improved calculations are clearly needed.

We should not forget that the arguments we use only give us upper limits for the scale when new physics phenomena must set in.
They can indeed set in significantly earlier, as was the case with the Higgs mass which is way below its unitarity limit.
In any case even if we take the largest of the numbers above, the given scales are many orders of magnitude smaller than any other 
indication of a breakdown in the SM such as Landau poles, the Planck scale or the scale at which the Higgs potential becomes metastable.

Let us now turn to the question what this implies for observation. As already mentioned the crucial energy scale is quite low possibly even in the reach of future circular hadron colliders. This opens exciting new possibilities.
The next question is, of course the size of the new effects. If the observational limits from the $Z$-peak or the cosmic rays are even close to being saturated cross sections for new phenomena are very large and would lead to quite spectacular effects. Indeed the cross sections
would then be so large that for a hadron collider such as the FCC, the suppression of high energy events by the fall in the particle distribution functions could be overcome and one could essentially utilise the full energy of the protons in each collision. 
On the other hand at the present state one cannot exclude the possibility that the SM is repaired ``in secret'': non-perturbative effects start to suppress the factorial growth of cross sections at or around the point when cross sections are minimal. If this is the only place where such repairs set in, effects would be unobservably small.  Yet, it seems unlikely that such repairs only happen in a place we can't observe. 
To clarify this further studies are needed to determine if and how non-perturbative effects can cure the factorial growth of cross sections and what observable facts are associated with it. 

Let us now engage in a bit of speculation. It has already been noted~\cite{Voloshin:1992xb,Argyres:1993xa,Smith:1993hz} that for certain special values of the masses tree-level threshold amplitudes for physical high multiplicity $2\to n$ processes vanish. For example for $VV\to nH$ one of these special values is $\kappa=m_{V}/m_{h}=g/(2\sqrt{2\lambda})=1/\sqrt{2}=0.71$. At the electroweak scale $\kappa=0.65$ for $m_{V}=m_{W}\approx 80$~GeV or using $m_{V}=m_{Z}\approx 91$~GeV it is $0.73$. This is close but still somewhat off the magic value $1/\sqrt{2}$.
Renormalisation group evolving the couplings to higher energy scales $\kappa$ grows and starting from the $W$-boson value $0.65$ it reaches $0.71$ well before the perturbative breakdown. This could hint towards a potential repair mechanism\footnote{It should be noted that in a diagram describing a single multiparticle process one would expect the coupling constants to be evaluated at different scales depending on the flow of energy and momentum through the respective vertex. So effectively one does not have a single value of $\kappa$ in any given diagram.} arising from special values of the coupling constants (cf. also~\cite{Voloshin:1992xb,Argyres:1993xa,Smith:1993hz}). At the same time one would still need a qualitative change compared to the standard perturbative behaviour
since $\kappa$ passes through the magic value quite quickly and factorial growth would resume and soon overpower everything.  

Perhaps even more speculatively one can also surmise that the rising cross sections indicate a more fundamental defect in the gauge-Higgs sector indicating the need for physics beyond the Standard Model at a quite low scale. 
Of course the growth in cross sections and amplitudes could simply be the usual behaviour of asymptotic series~\cite{Dyson:1952tj,Lipatov:1976ny,Boyd} that should only be used until a finite 
order and then be replaced by a different approximation scheme.
Looking only at the spectral representation of the full propagator this could indeed be what one would expect, since the multiparticle amplitudes correspond to higher and higher order corrections in the perturbative calculation. Nevertheless let us note that for real high multiplicity processes the tree-level amplitudes we have considered, correspond to the leading order behaviour.
It is also interesting to compare to another system where high multiplicity amplitudes have been calculated, unbroken non-abelian gauge theories. In these systems the leading order behaviour in the number of colors features a cancellation such that the high multiplicity minimal helicity violating amplitudes do not grow factorially~\cite{Berends:1987me,Mangano:1990by}. This is in stark contrast to the gauge-Higgs system. This could be a hint to a crucial difference between the two systems and a potential deep problem in the gauge Higgs system.

\section*{Acknowledgements}
We would like to thank Martin Bauer, Tilman Plehn and Gavin Salam for interesting discussions. JJ~ thanks IPPP for hospitality, the research of VVK is supported by the STFC, the Royal Society and the Wolfson foundation.

\end{document}